\documentclass[12pt]{article}
\usepackage{amsmath}
\usepackage{graphicx} 
\usepackage{natbib}
\usepackage{url}  

\newcommand{\blind}{0}
\usepackage{algorithmic}
\usepackage{graphicx,psfrag,epsf}
\usepackage{enumerate}
\usepackage{natbib}
\usepackage{tabularx} 
\usepackage{lscape} 
\usepackage{url} 
\usepackage{enumitem} 
\usepackage{multirow}
\usepackage{accents}
\usepackage{graphicx}
\usepackage{rotating}
\usepackage{subcaption}
\usepackage{mwe}
\usepackage{float}
\usepackage{booktabs,makecell}
\usepackage{amsthm,amssymb}
\usepackage{amsfonts} 
\usepackage{dsfont}
\usepackage{amsmath}
\usepackage{amssymb}
\usepackage{setspace}
\usepackage{hyperref}
\usepackage{multirow}
\usepackage{booktabs}
\usepackage{bbm}
\usepackage{mathrsfs} 
\usepackage{bbm}
\usepackage{booktabs}
\usepackage{tcolorbox}
\usepackage{multirow}
\usepackage{rotating}
\hypersetup{
     colorlinks   = true,
  linkcolor=blue,
            urlcolor=blue,
            citecolor=blue
}
\usepackage{xr}
\makeatletter
\newcommand*{\addFileDependency}[1]{
  \typeout{(#1)}
  \@addtofilelist{#1}
  \IfFileExists{#1}{}{\typeout{No file #1.}}
}
\makeatother

\newcommand*{\myexternaldocument}[1]{%
    \externaldocument{#1}%
    \addFileDependency{#1.tex}%
    \addFileDependency{#1.aux}%
}

\myexternaldocument{sup}
\usepackage[ruled,vlined,onelanguage]{algorithm2e}
\usepackage{pdflscape}
\usepackage{fancyhdr}

\fancypagestyle{mylandscape}{
\fancyhf{} 
\fancyfoot{
\makebox[\textwidth][r]{
  \rlap{\hspace{.75cm}
    \smash{
      \raisebox{4.87in}{
        \rotatebox{90}{\thepage}}}}}}
}

\usepackage[left, pagewise]{lineno}  

\begin{document}

\def\spacingset#1{\renewcommand{\baselinestretch}%
{#1}\small\normalsize} \spacingset{1}


\def\bmY{\boldsymbol{Y}}
\def\bmW{\boldsymbol{W}}
\def\bmM{\boldsymbol{M}}
\def\bmE{\boldsymbol{E}}
\def\bmL{\boldsymbol{L}}
\def\bmlambda{\boldsymbol{\lambda}}
\def\bmrho{\boldsymbol{\rho}}
\def\Cov{{\rm Cov}}
\def\Avar{{\rm Avar}}
\def\rmvec{{\rm vec}}
\def\bmSigma{\boldsymbol{\boldsymbol{\Sigma}}}
\def\mbR{\mathbb{R}}
\def\mbW{\mathbb{W}}
\def\bmI{\boldsymbol{I}}
\def\bmy{\boldsymbol{y}}
\def\bmvarepsilon{\boldsymbol{\varepsilon}}
\def\bmxi{\boldsymbol{\xi}}
\def\bmtheta{\boldsymbol{\theta}}
\def\hbmtheta{\boldsymbol{\widehat\theta}}
\def\tbmtheta{\boldsymbol{\widetilde\theta}}
\def\bbmtheta{\boldsymbol{\bar\theta}}
\def\mL{\mathscr{L}}
\def\tmL{\widetilde {\mL}}
\def\bmeta{{\boldsymbol{\eta}}}
\def\lbk{\left\{}
\def\rbk{\right\}}
\def\lak{\left|}
\def\rak{\right|}
\def\lmk{\left[}
\def\rmk{\right]}
\def\lsk{\left(}
\def\rsk{\right)}
\def\bmx{\boldsymbol{x}}
\def\bmz{\boldsymbol{z}}
\def\bmv{\boldsymbol{v}}
\def\bmmu{\boldsymbol{\mu}}
\def\bmzeta{\bmzeta}
\def\bmsigma{\boldsymbol{\sigma}}
\def\bmtheta{{\boldsymbol{\theta}}}
\def\bml{\boldsymbol{l}}
\def\bmo{\boldsymbol{0}}
\def\diag{{\rm diag}}
\def\bmrho{\boldsymbol{\rho}}
\def\E{{\rm E}}
\def\bmgamma{\boldsymbol{\gamma}}
\def\bmalpha{\boldsymbol{\alpha}}
\def\bmvtheta{\widehat{\bmtheta}}
\def\laak{\left\|}
\def\raak{\right\|}
\def\OLS{{\rm OLS}}
\def\GLS{{\rm GLS}}
\def\FGLS{{\rm FGLS}}
\def\LS{{\rm LS}}
\def\FLS{{\rm FLS}}
\def\tr{{\rm tr}}
\def\mbY{\mathbb{Y}}
\def\mbX{\mathbb{X}}
\def\rmE{{\rm E}}
\def\rmI{{\rm I}}
\def\bmI{{\boldsymbol{I}}}
\def\rmm{{\rm m}}
\def\rmP{{\rm P}}
\def\Cov{{\rm Cov}}
\def\var{{\rm Var}}
\def\mS{{\mathcal{S}}}
\def\mN{{\mathcal{N}}}
\def\rmU{{\mathrm{U}}}
\def\rmV{{\mathrm{V}}}
\def\rmQ{{\mathrm{Q}}}
\def\bmU{\boldsymbol{U}}
\def\bmV{{\boldsymbol{V}}}
\def\bmQ{{\boldsymbol{Q}}}
\def\bmH{{\boldsymbol{H}}}
\def\sgn{\operatorname{sgn}}
\def\bu{\boldsymbol u}
\def\mumed{\mu_{\text{med}}}
\def\vech{\mathrm{vech}}
\def\vec{\mathrm{vec}}
\def\bmZ{\boldsymbol{Z}}
\def\bmG{\boldsymbol{G}}
\def\bmX{\boldsymbol{X}}
\def\bmS{\boldsymbol{S}}
\def\bmh{\boldsymbol{h}}
\def\bmu{\boldsymbol{u}}
\def\bmzeta{\boldsymbol{\zeta}}
\def\bmt{\boldsymbol{t}}
\def\bmcalV{\boldsymbol{\mathcal{V}}}
\def\bms{\boldsymbol{s}}
\def\bmq{\boldsymbol{q}}
\def\bmk{\boldsymbol{k}}
\def\bmcalH{\boldsymbol{\mathcal{H}}}
\def\bmg{\boldsymbol{g}}
\def\bmw{\boldsymbol{w}}
\def\bmTheta{\boldsymbol{\Theta}}
\def\bmcalA{\boldsymbol{\mathcal{A}}}
\def\bmcalB{\boldsymbol{\mathcal{B}}}
\def\bmrmB{\boldsymbol{\mathrm{B}}}
\def\bmrmV{\boldsymbol{\mathrm{V}}}
\def\bmDelta{\boldsymbol{\Delta}}
\def\for{\textbf{\text{for }}}
\def\do{\textbf{\text{do }}}
\def\endfor{\textbf{\text{end for }}}
\def\mG{\mathcal{G}}
\def\bmcalG{\boldsymbol{\mathcal{G}}}

\def\be{\begin{equation}}
\def\ee{\end{equation}} 
\def\ben{\begin{equation*}}
\def\een{\end{equation*}}
\def\bea{\begin{eqnarray}}
\def\eea{\end{eqnarray}}
\def\bda{\begin{eqnarray*}}
\def\eda{\end{eqnarray*}}
\numberwithin{equation}{section}
\def\m{\color{magenta}}
\def\b{\color{blue}}
\def\red{\color{red}}
\def\here{{\color{magenta}Tao is up to here.}}
\def\bmY{\boldsymbol{Y}}
\def\bmX{\boldsymbol{X}}
\def\bmT{\boldsymbol{T}}
\def\bmA{\boldsymbol{A}}
\def\bmPsi{{\boldsymbol{\Psi}}}
\def\bmpsi{{\boldsymbol{\psi}}}
\def\bmUpsilon{\boldsymbol{\Upsilon}}
\def\bmcalZ{\boldsymbol{\mathcal{Z}}}
\def\bmrmZ{\boldsymbol{\mathrm{Z}}}
\def\bmbbE{\boldsymbol{\mathbb{E}}}
\def\bmmP{\boldsymbol{\mathcal{P}}}
\def\bmmH{\boldsymbol{\mathcal{H}}}
\def\bmvartheta{\boldsymbol{\bmtheta}}
\def\bmnu{\boldsymbol{\nu}}
\def\bmvarsigma{\boldsymbol{\varsigma}}
\def\bmvarSigma{\boldsymbol{\varSigma}}
\def\bmD{\boldsymbol{D}}

\newcommand{\Grad}{\nabla\!\!\!\!\nabla}
\newtheorem{theorem}{Theorem}
\newtheorem{remark}{Remark}
\newtheorem{proposition}{Proposition}
\newtheorem{corollary}{Corollary}
\newtheorem{assumption}{Assumption}
\newtheorem{lemma}{Lemma}
\newtheorem{definition}{Definition}
\newtheorem{example}{Example}

\definecolor{darkgreen}{RGB}{0, 176, 80}
\definecolor{islrorange}{RGB}{205, 100, 32}
\definecolor{islrblue}{RGB}{93, 165, 216}
\definecolor{whaleblue}{RGB}{51, 51, 178}
\definecolor{highlightblue}{RGB}{68, 68, 255}
\definecolor{pptorange}{RGB}{237, 125, 49}
\definecolor{pptgreen}{RGB}{112, 173, 71}
\definecolor{pptpurple}{RGB}{114, 51, 162}
\definecolor{highlightred}{RGB}{204, 0, 0}
  
\newcommand{\song}{\CJKfamily{song}}
\newcommand{\fs}{\CJKfamily{fs}}
\newcommand*{\hl}[1]{\textcolor{highlightred}{#1}}
\newcommand*{\pl}[1]{\textcolor{pptpurple}{#1}}


\if0\blind
{\spacingset{1.5} 
  \title{\bf Regularization and Selection in A Directed Network Model with Nodal Homophily and Nodal Effects}
  \date{}
  \author{
    Zhaoyu Xing \\ \textit{Xiamen University, China} 
    \and 
    Y. X. Rachel Wang \\ \textit{University of Sydney, Australia}
    \and 
    Andrew T. A. Wood \\ \textit{Australian National University, Australia}
    \and 
    Tao Zou$^*$ \\ \textit{Australian National University, Australia}
  }
  \maketitle
} \fi

\if1\blind
{\spacingset{2} 
  \bigskip
  \bigskip
  \bigskip
  \begin{center}
    {\LARGE\bf [Title Hidden for Blind Review]}
  \end{center}
  \medskip
} \fi

\bigskip 
\begin{abstract} 
{
This article introduces a regularization and selection methods for directed networks with nodal homophily and nodal effects. The proposed approach not only preserves the statistical efficiency of the resulting estimator, but also ensures that the selection of nodal homophily and nodal effects is scalable with large-scale network data and multiple nodal features. 
In particular, we propose a directed random network model with nodal homophily and nodal effects, which includes the nodal features in the probability density of random networks. Subsequently, we propose a regularized maximum likelihood estimator with an adaptive LASSO-type regularizer. We demonstrate that the regularized estimator exhibits the consistency and possesses the oracle properties. In addition, we propose a network Bayesian information criterion which ensures the selection consistency while tuning the model.
Simulation experiments are conducted to demonstrate the excellent numerical performance. An online friendship network among musicians with nodal musical preference is used to illustrate the usefulness of the proposed new network model in network-related empirical analysis. 
}
\end{abstract}

\noindent%
\vfill

\newpage
\spacingset{1.9} 

\section{Introduction} 

Complex networks are ubiquitous across various scientific disciplines and garnered sustained attentions from scholars. Examples include research on friendship networks in social sciences, studies on game-theoretic networks in economics, and investigations into brain networks in biology. As network science fundamentally emphasizes the characterization of interactive relationships among distinct entities, most of the classical statistical methods and probability distributions are not appropriate to model network data. Concurrently, rapid development in scientific technologies have enabled the widespread collection of large-scale networks with nodal features, which brings challenges in developing robust and scalable approach to model the complex networks with multiple nodal features.    


To model the random networks, several edge-based random network models are proposed in early years, such as Erdős-Rényi (ER) random network model \citep{ERnetwork1959} where each edges are connected randomly with a specific probability, Barab\'{a}si--Albert (BA) network \citep{albert2002statistical} and small-world networks \citep{Watts1998}. Although these network models with specific generating mechanism are commonly used to generate networks in applications\citep{Zheng2006JASA, DePaula2020, Neal2003, XingCalms2025}, they have almost no power to fit complex networks with statistical inference. 
Stochastic block model \citep{Zhao2012, LeiRinaldo} considers the random networks with grouped nodes and independent edges from Bernoulli distribution. Stochastic block model and its modifications can characterize the difference in edge probability between clusters, but they include only the pure network structure and can hardly handle the information from nodal features. 
To model more complex randomness in networks while considering nodal information, dyadic models \citep{zheng2006many, hoff2005bilinear} consider to formulate the probabilities of ``social ties" directly in the applications of econometrics and social science. However, these methods do not forces on nodal homophily and nodal effects. \citet{graham2017econometric, Yan2016AOS} use the distances computed from all nodal features, which limits their ability to separate nodal homophily and nodal effects from individual effects or to select informative nodal features.  


Recently, massive and multivariate nodal homophily and nodal effects are widely considered in empirical analysis of networks \citep{McPherson2001, Toivonen2009}. Numerous studies on social networks, where the vertices represent individuals or volunteers, have shown that gender-based homophily significantly affects the formation of network structures \citep{Hunter2007, Stewart2019}. Other significant nodal effects have also been identified in both experimental and survey-based network data, including homophily and individual effects related to race, degree, and smoking behavior \citep{Wimmer2010, McMillan2022, Wangetal2024}. However, these empirical analyses typically present only one fitted model with a specific set of nodal features. Although the detected homophily patterns can often be well explained in context, we are concerned about the robustness of such conclusions across different network samples, as well as the scalability of these methods when applied to large-scale networks with complex and multivariate nodal structures.

When modeling networks with a large number of nodal features, it is crucial to identify the true nodal homophily and nodal effects from among many candidate variables. For example, in the classic analysis of online social networks, each user account typically includes a rich set of features in addition to the network of relationships \citep{SK2015dataset, NEO2016}. It is natural and intuitive to consider attributes like gender and race when analyzing social interactions, but selecting relevant homophily effects becomes challenging in general settings without prior knowledge. In particular, when focusing on directed social networks among musicians, which is further discussed in Section \ref{S5}, preferences for different music genres may play a central role. To identify the genres that strongly influence social interactions and to quantify their effects on the formation of friendships, we require a tailored directed random network model that accommodates sparse nodal homophily and individual nodal effects.
Another hot topic in network analysis is the investigation of factors driving academic collaboration among scholars, departments, and institutions. This is typically studied using co-authorship and citation networks across disciplines, such as high-dimensional statistics \citep{gao2024citation} and energy physics \citep{lehmann2003citation, newman2001scientific}. These networks also involve multiple informative nodal attributes - such as scholars’ educational backgrounds, research interests, and publication keywords - that should be incorporated into the analysis. To detect and select the most influential nodal homophily and nodal effects in these contexts, a specialized random network model is again necessary. 

Motivated by these practical needs in empirical network analysis, we propose a novel directed network model that incorporates both nodal homophily and nodal effects. The model is designed to efficiently handle large-scale networks with multiple nodal features and to consistently identify the true subset of relevant nodal homophily and effects through a data-driven approach. Specifically, we introduce a new probability model for random directed networks that includes dyadic reciprocity, nodal homophily and nodal effects. We develop a regularized maximum likelihood estimator with an adaptive regularizer to estimate parameters and complete the data-driven selection of nodal homophily and nodal effects. We prove the proposed regularized estimator is consistent and possesses the oracle properties, which indicates that the parameters of relevant nodal homophily and nodal effects can be estimated as efficiently as if the true subset were known as the prior knowledge. Furthermore, we propose a network Bayesian information criterion (NBIC) for selecting the tuning parameter, and prove that it ensures consistent model selection. 


The rest of the article is organized as follows.  Section \ref{S2} introduces the directed network model to capture the nodal homophily and nodal effects. Section \ref{S3} presents the theoretical properties of our proposed method. Simulation studies are provided in Section \ref{S4}, and a real data analysis to demonstrate the usefulness of our method is given in Section \ref{S5}.  Section \ref{S6} concludes the paper with a discussion. Technical conditions and assumptions are given in the Appendix. Proofs of the main theorems, supporting lemmas, and additional numerical results are provided in the supplementary material.

\section{A Directed Network Model with Nodal Homophily and Nodal Effects}
\label{S2} 

Let $\mathcal{G} = \{\mathcal{N},\mathcal{E}\}$ be a directed binary graph with node set $\mathcal{N}=[n]$ and edge set $\mathcal{E}$, where $[n]=\{1,\cdots, n\}$. We exclude the self-loops in networks, so the total number of potential directed edges is $n(n-1)$. Let $X=(X_{j_1j_2})_{j_1,j_2\in[n]} \in \{0,1\}^{n\times n}$ denote the adjacency matrix of binary 
network $\mathcal{G}$, where $\diag(X)=(X_{11},\cdots,X_{nn})^\top = 0$. 
Suppose that each node $j\in[n]$ is associated with $m$ groups of nodal feature vectors, denoted by $v_{jk}$, $k=1,\cdots,m$. For each group $k$, the feature vector $v_{jk}$ has dimension $r_k\geq 1$. We stack those feature vectors of node $j$ into a single vector $v_j=(v_{j1}^ \top,\cdots, v_{jm}^ \top)^ \top \in \mbR ^r$, where $r=\sum_{k=1}^m r_k $. Let $V =  (v_{1},\cdots,v_{n})^\top \in \mathbb{R}^{n \times r}$ denote the resulting nodal feature matrix. Examples of such group structures of nodal features can be found in \citet{Yan2019JASA,bramoulle2020peer}. If no prior grouping information is available, we set $r_1 = \cdots = r_m = 1$ so that $r = m$. The nodal features may be a mix of categorical and continuous variables. Without loss of generality, we assume that all categorical features are coded as dummy variables.
Let $x = (x_{j_1j_2})_{j_1,j_2 \in [n]} \in {0,1}^{n \times n}$ be a realization of the random graph $X$, where $\operatorname{diag}(x) = 0$. We assume that $X$ follows a probability distribution parameterized by $\theta_1$, $\theta_2$, 
$\gamma=(\gamma_1,\cdots,\gamma_m)$, 
$\zeta=(\zeta_1,\cdots,\zeta_m)$ and $\eta=(\eta_1,\cdots,\eta_m)$ as follows: 
\bea
&& \rmP( X = x |\theta_1,\theta_2,\gamma,\zeta,\eta ; V )\nonumber\\ &=&\frac{1}{z( \theta_1,\theta_2,\gamma,\zeta,\eta; V )} \exp \Big\{\theta_1\sum_{j_1,j_2\in[n];j_1<j_2} x_{j_1j_2}x_{j_2j_1} + \theta_2 \sum_{j_1,j_2\in[n];j_1 \neq j_2} x_{j_1j_2}  \nonumber\\    
& & + \sum_{k=1}^m \gamma_k s_k^{\textrm{H}}(x,V)  + \sum_{k=1}^m \zeta_k s_k^{\textrm{O}}(x,V)+ \sum_{k=1}^m \eta_k s_k^{\textrm{I}}(x,V) \Big\}
\label{S2:eq:PDF}
\eea 
where the first term $\sum_{j_1,j_2\in[n];j_1<j_2} x_{j_1j_2}x_{j_2j_1}$ represents the sample measurement of reciprocity, the second term
$\sum_{j_1,j_2\in[n];j_1<j_2} x_{j_1j_2}$ is the total count of edge. The nodal homophily from $k$-th group of nodal features is given by $s_{k}^{\textrm{H}}(x,V)=\sum_{j_1,j_2\in[n];j_1 < j_2} x_{j_1j_2} w(v_{j_1k},v_{j_2k})$, 
$s_{k}^{\textrm{O}}(x,V)=\sum_{j_1,j_2\in[n];j_1\neq j_2} x_{j_1j_2} g(v_{j_1k})$ and $s_{k}^{\textrm{I}}(x,V)=\sum_{j_1,j_2\in[n];j_1\neq j_2} x_{j_1j_2} g(v_{j_2k})$ represent the in-edge and out-edge effect from $k$-th group of nodal features, respectively. The normalizing constant $z( \theta_1, \theta_2,  \gamma,\zeta,\eta;V )$ is chosen such that
\[
\sum_{x \in \{0,1\}^{n\times n}\textrm{with $\diag( x )=0$}} \rmP( X = x |\theta_1,\theta_2,\gamma,\zeta,\eta)=1.
\]

In the above density \eqref{S2:eq:PDF}, $w(v_{j_1 k},v_{j_2 k})$ denotes the similarity measure between variables $v_{j_1 k}$ and $v_{j_2 k}$. We adapt the concept of pairwise comparisons \citep{johnson1992applied} to gauge the similarity between these variables. For continuous $v_{j k}$, a commonly used similarity measure is the Euclidean distance $\|v_{j_1 k}-v_{j_2 k}\|_2$, which can be further extended by defining similarity measures as decreasing functions of distance. For instance, we can use $w(v_{j_1 k},v_{j_2 k})=\exp \left\{-\|v_{j_1 k}-v_{j_2 k}\|_2^2\right\}$ as seen in related network studies \citep{Zou02012017, chen2010parallel, xing2024golfs}.
For discrete $v_{j k}$, similarity measures such as the Hamming distance can be applied \citep{johnson1992applied}. Specifically, $w(v_{j_1 k},v_{j_2 k})=1$ if all dimensions of $v_{j_1 k}$ and $v_{j_2 k}$ are in the same category, and $w(v_{j_1 k},v_{j_2 k})=0$ otherwise. 
Since the distance and similarity measures are based on nodal features, the nodal homophily and nodal effects included in our model could be more general including the combinations of nodal features. Consequently, the selection of nodal homophily and nodal effects is more flexible than merely selecting nodal features. We show that some commonly used models in closely related empirical analyses can be viewed as special cases of our model in \eqref{S2:eq:PDF} as outlined below.

\begin{example}[Erdős-Rényi (ER) model]  
\label{ER} 
The well-known ER random network model \citep{ERnetwork1959} are commonly used as the benchmark model in network modeling, where the probability for each edge is 
\[
\rmP( X_{j_1j_2} = 1 ) = \wp, ~ j_1,j_2 \in [n], j_1\neq j_2,
\]
and all the edges are generated independently. In Section \ref{sup:S1} of the supplementary material, we have shown that the ER model is a special case of \eqref{S2:eq:PDF} with $\theta_1=\gamma=\zeta=\eta=0$. 
\end{example}

\begin{example}[The exponential random graph models (ERGMs) with nodal homophily]  
\label{Example:ERGM_H}       
ERGMs are commonly used in empirical analysis to quantify the nodal homophily effects \citep{Wimmer2010, McMillan2022, Wangetal2024}.
The probability for one edge could be written as
\[
\label{s2:ERGM_H}
\rmP( X_{j_1j_2} = 1 ) \propto \exp \{ \theta_2 + \sum_{k=1}^m \gamma_k w(v_{j_1 k}, v_{j_2 k}) \} ~ j_1,j_2 \in [n], j_1\neq j_2,
\]
where $w(\cdot)$ are constructed based on the similarity measure between a single nodal feature \citep{hunter2008ergm}. In Section \ref{sup:S1} of the supplementary material, we have shown that the ERGMs with nodal homophily are special cases of \eqref{S2:eq:PDF} with $r=m$ and $\theta_1=\zeta = \eta=0$. Based on (\ref{s2:ERGM_H}), the probability of the appearance of one edge depends on the overall level of network density controlled by $\theta_1$ and the nodal homophily determined by $\gamma$. The sparsity of the $\gamma$ indicates that only few of the homophily from nodal features have significant effects on the edge probability. 
\end{example}

\begin{example}[The exponential random graph models (ERGM) with nodal effects] 
\label{s2:Example:ERGM_NE} 
Besides nodal homophily, the nodal effects are also commonly considered in the ERGMs for directed networks \citep{jiang2015natergm} where the edge probability is 
\[
\label{s2:eq:ERGM_NE}
\rmP( X_{j_1j_2} = 1 ) \propto \exp \{ \theta_2 + \sum_{k=1}^m \zeta_k g(v_{j_1k}) + \sum_{k=1}^m \eta_k g(v_{j_2k})  \} ~ j_1,j_2 \in [n], j_1\neq j_2.
\]
In Section \ref{sup:S1} of the supplementary material, we have shown that the ERGMs with nodal nodal effects are special cases of \eqref{S2:eq:PDF} with $r=m$ and $\theta_1=\gamma=0$. 
Based on (\ref{s2:eq:ERGM_NE}), probability of one edge depends on the overall density characterized by $\theta_2$, the out-edge nodal effects and in-edge nodal effects controlled by $\zeta$ and $\eta$, respectively. 
The sparsity of the $\zeta$ and $\eta$ indicate that only few of nodal effects are influential on the edge probability.  
\end{example}


Based on the probability density for networks defined in \eqref{S2:eq:PDF}, we further focus on the dyadic relationships in directed random networks. There are four potential dyads in binary networks: $(1,1)$, $(1,0)$, $(0,1)$, and $(0,0)$. The probability of a specific dyad is determined by the overall density, the effect of reciprocity, nodal homophily, and nodal effects. The magnitude of the marginal effects is controlled by the corresponding parameters. Without loss of generality, we use the empty dyad $(0,0)$ as the reference group. Then we obtain the probability of one dyad with one directed edge from node $j_1$ to node $j_2$ and no reciprocity as 
\begin{equation}   
\label{s2:eq:D10}
\rmP \{D_{j_1 j_2}=(1,0)\} \propto \exp \big\{ \theta_2 + \sum_{k=1}^m \gamma_k w(v_{j_1k},v_{j_2k}) + \sum_{k=1}^m \zeta_k g(v_{j_1k}) + 
\sum_{k=1}^m \eta_k g(v_{j_2k}) \big\} := T_{10}
\end{equation}
where $D_{j_1 j_2}=(X_{j_1j_2}, X_{j_2j_1})$ denote one dyad of the network between node $j_1$ and $j_2$, and we denote the corresponding realization as $d_{j_1 j_2}=(x_{j_1j_2}, x_{j_2j_1})$. In equation (\ref{s2:eq:D10}), $\theta_2$ serves as the intercept that characterizes the overall network density, which is equivalent to the parameter $\wp$ in Example \ref{ER} when the other parameters are all zero. The coefficient $\gamma_k$ captures the marginal effect from nodal homophily between two nodes, $\zeta_k$ and $\eta_k$ quantify the nodal effects from $k$-th group of nodal features $v_k$ of the origin and destination.  The sparsity in $\gamma$, $\zeta$, and $\eta$ reflects the fact that only a few influential nodal features from $V$ affect the network generation process. The sparsity of effects from multiple features is widely discussed and accepted as an approach to real-world situations \citep{LiZhong2012feature, fan2008sure, hastie2015statistical}.

Similarly, we have the probability for the dyad with inverse directed edge node $j_2$ to node $j_1$ and no reciprocity as 
\begin{equation}
 \rmP \{D_{j_1 j_2}=(0,1)\} \propto \exp \big\{ \theta_2 + \sum_{k=1}^m \gamma_k w(v_{j_1k},v_{j_2k}) + \sum_{k=1}^m \zeta_k g(v_{j_2k}) + 
  \sum_{k=1}^m \eta_k g(v_{j_1k})   \big\} :=  T_{01}
    \label{s2:eq:D01}
\end{equation}
and the probability for one dyad with two edges between node $j_1$ and node $j_2$ as 
\begin{align} 
\rmP \{D_{j_1 j_2} = (1,1)\} ~\propto ~ & \exp \big\{ \theta_1 + \theta_2 \times (1+1) \nonumber \\ 
    & + \sum_{k=1}^m \zeta_k \{g(v_{j_1k}) + g(v_{j_2k})\} + 
  \sum_{k=1}^m \eta_k \{g(v_{j_1k})+g(v_{j_2k})\} \nonumber\\
    & +\sum_{k=1}^m \gamma_k \times 2w(v_{j_1k},v_{j_2k})   \big\} := T_{11}
    \label{s2:eq:D11}
\end{align}
where $\theta_1$ is the parameter captures the force of reciprocation in one dyad. If $\theta_1 = 0$, then there is no additional tendency for reciprocity in the random network, given the nodal heterogeneity in homophily and nodal effects included in our model. Although there is no direct dependence between the dyadic pairs $x_{j_1 j_2}$ and $x_{j_2 j_1}$, the distributions of dyads are not identical due to the influence of nodal features.

For simplicity, we rewrite our model as 
\bea 
\rmP(X=x|\theta) &=& \frac{1}{z( \theta)}\exp\Big\{ 
\theta_1 \sum_{ j_1,j_2\in[n]; j_1<j_2 } x_{j_1 j_2}x_{j_2 j_1}  +  \theta_2 \sum_{ j_1,j_2\in[n]; j_1\neq j_2 } x_{j_1 j_2} + \nonumber \\ 
&& \sum_{k=1}^m \theta_{2+k} s_k^{\textrm{H}}(x,V)  + \sum_{k=1}^m \theta_{2+m+k} s_k^{\textrm{O}}(x,V) + 
\sum_{k=1}^m \theta_{2+2m+k} s_k^{\textrm{I}}(x,V)  \Big\} \\  
&:=& \frac{\exp\{ \theta ^\top s ( x , V )\}}{z(\theta)}
\label{S2:eq:idtPDF} 
\eea 
with normalizing constant $z(\theta)=T_{10} +T_{01} +T_{11} +1$, the $p$-dimensional unknown parameter vector $\theta=(\theta_1,\cdots,\theta_{2+3m})$ where $p=3m+2$ and  
 $(\theta_{3},\cdots,\theta_{2+m})=(\gamma_1,\cdots,\gamma_m)$,  $(\theta_{3+m},\cdots,\theta_{2+2m}) = (\zeta_1,\cdots,\zeta_m),$
$(\theta_{3+2m},\cdots,\theta_{2+3m}) = (\eta_1,\cdots,\eta_m)$ 
with the corresponding sufficient statistics $s(x,V) =\{s_1(x,V), \cdots,s_{2+3m}(x,V)\}$ where
\begin{center}
$\{s_{3}(x,V),\cdots,s_{2+m}(x,V)\} = \{s_1^\textrm{H}(x,V),\cdots,s_{m}^\textrm{H}(x,V)\}$,\\
$\{s_{3+m}(x,V),\cdots,s_{2+2m}(x,V)\} = \{s_1^\textrm{O}(x,V),\cdots,s_{m}^\textrm{O}(x,V)\}
$, \\
$\{s_{3+2m}(x,V),\cdots,s_{2+3m}(x,V)\} = \{s_1^\textrm{I}(x,V),\cdots,s_{m}^\textrm{I}(x,V)\}$.
\end{center}

In the remainder of this paper, we sometimes denote $P( X = x ) \propto \exp\{ \theta ^\top s ( x , V )\}$  for simplicity, omitting the normalizing constant $z(\theta)$.

\section{Methodology}
\label{S3}

\subsection{Reparameterization and Regularized Maximum Likelihood Estimation}

In the dyadic model introduced in the previous section, there are only four possible observations for each dyad. By combining \eqref{s2:eq:D10}, \eqref{s2:eq:D01}, and \eqref{s2:eq:D11}, with some derivation, we can reparameterize the model using a one-hot encoding of dyads as follows:
\bea
\label{S3:multinominal:Prob}
\frac{P(y_{j_1,j_2}=c)}{P(y_{j_1,j_2}=4)}=\exp\{S_{j_1,j_2} ^\top \tilde{\theta}_c\}, c=1,2,3
\eea 
where $y\in \{1,2,3,4\}^{n(n-1)/2}$ is a categorical vector with four levels representing the dyads \{(1,1), (1,0), (0,1), (0,0)\} respectively, $\tilde{\theta}_c\in \mathbb{R}^{p\times 1}$ are the unknown parameters, and $\mathrm{S} \in\mathbb{R}^{N \times p}$ is the loading matrix with $N=n(n-1)/2$ dyads and $p=3m+2$ dimensions. We denote $x_{-(j_1,j_2)}$ as the observed networks $x$ excluding the dyad between the nodes $j_1$ and $j_2$. Then $S_{j_1,j_2} \in \mathbb{R}^{p}$ for the dyad is the vector characterizes the difference in network statistics $s(x,V) \in \mathbb{R}^p$ between network $\{ x_{-(j_1,j_2)} \cup (1,1)\}$ and $\{ x_{-(j_1,j_2)} \cup (0,0)\}$, which can be written as
$$S_{j_1,j_2}=(1,2,w(v_{j_1,1},v_{j_2,1}),\cdots,w(v_{j_1,k},v_{j_2,k}),g(v_{j_1,1}),\cdots,g(v_{j_1,k}),g(v_{j_2,1}),\cdots,g(v_{j_2,k})) ^\top.$$
Since the mutual edges in a dyad are affected by nodal features with the same parameters in the directed random network model \eqref{S2:eq:idtPDF}, and we use the empty dyad as the reference group, we can reparameterize the model as the following non-identical distribution for dyads:
\bea
P(y_{j_1,j_2}=c) = f_c(S_{j_1,j_2},\theta) \propto \exp\{S_{j_1,j_2}^\top \tilde{\delta}_c\} , c=1,2,3, 
\eea
where $\tilde{\delta}_c = A_c\theta, c=1,2,3$ and $A_c$ is a fixed diagonal matrix. Please refer to Section \ref{Sup:Lemma1proof} in the supplementary material for further details.

Assuming the proposed directed random network model in \eqref{S2:eq:idtPDF} is well-specified, the likelihood of an observed network $y$,  incorporating nodal homophily and nodal effects, is given by 
\bea 
\mathcal{L}(\theta) &=& \log \left\{  \prod_{j_1,j_2 \in[n];j_1 < j_2} \rmP \left( D_{j_1j_2}=(x_{j_1j_2},x_{j_2j_1})|\theta \right) \right\}  \nonumber  \\ 
&=& \sum_{j_1,j_2 \in[n];j_1 < j_2} \sum_{c=1}^4 y_{j_1,j_2,c} \log f_c(S_{j_1,j_2},\theta)
\label{S3:Likelihood}  
\eea  
where $f_c(S_{j_1,j_2},\theta)$ denotes the probability mass function for the dyad between nodes $j_1$ and $j_2$, corresponding to category $c=\{1,2,3,4\}$.  The dyadic distribution is not identical across all pairs due to the heterogeneity in nodal features $S_{j_1,j_2}$. Here $y_{j_1,j_2,c}$ is the one-hot indicator for the observed category of the dyad $D_{j_1,j_2}$.  

We propose a regularized maximum-likelihood estimation (RMLE) for our directed random network model in \eqref{S2:eq:idtPDF} as
\begin{equation}
\label{S2:eq:RMLE}
\hat{\theta}_{RMLE} = \arg \min_{\theta\in\Theta} 
\left\{ -\mathcal{L}(\theta)  +  \lambda_N \sum_{k =1}^p w_k \left| \theta_{k}\right|_1 \right\}       
\end{equation}
where $\lambda_N$ is a tuning parameter, $\theta_{k}$ denotes the $k$-th element of vector $\theta$, and $w=(w_1,\cdots,w_p)^\top$ is a pre-specified adaptive weight vector that can be selected via data-driven methods.
In practice, the adaptive weight $w_k$ can be constructed as $1 / |\tilde{\theta}_k| ^ \gamma$ for some $\gamma>0$, where $\tilde{\theta}=(\tilde{\theta}_1,\cdots,\tilde{\theta}_p)$ is a $\sqrt{N}$-consistent estimator for $\theta$. According to Lemma \ref{lemma1} in the Supplementary Material, a natural choice for $\tilde{\theta}$ is the non-regularized maximum likelihood estimator $\hat{\theta}_{MLE} = \arg\max_{\theta\in\Theta} \mathcal{L}(\theta)$. In our simulations in Section \ref{S4}, we follow the recommendation by \citet{zou2006adaptive} and set $\gamma=1$.

Let $\theta_0=(\theta_{1,0},\cdots,\theta_{p,0})^\top$ denote the true parameter vector of $\theta$. As shown in Lemma \ref{lemma1} of the Supplementary Material, he non-regularized maximum likelihood estimator $\hat{\theta}_{MLE}$ is $\sqrt{N}$-consistent to $\theta_0$. Consequently, for any $k\in[p]$, if $\theta_{0,k}=0$, then $w_k=1 / |\hat{\theta}_{MLE,k}|$ will goes to infinity in probability as the size of network increases. Otherwise, if $\theta_{0,k}\neq0$, $w_k$ converges to a finite positive constant in probability.
This asymptotic property of the adaptive weights plays a critical role in variable selection. Specifically, as the network size increases, the divergence of $w_k$ for zero coefficients induces an increasingly large penalty in the regularization term, which drives the corresponding estimates toward zero. In contrast, for nonzero coefficients, the bounded $w_k$ ensures that the penalty remains moderate for accurate estimation. ensures that the penalty remains moderate, allowing accurate estimation. Consequently, the proposed regularized estimator $\hat{\theta}$ in \eqref{S2:eq:RMLE} achieves both consistent parameter estimation and consistent selection of relevant nodal homophily and nodal effects. In the following section, we establish that the regularize maximum likelihood estimator $\hat{\theta}$ in \eqref{S2:eq:RMLE} enjoys $\sqrt{N}$-consistency and the oracle properties. We further introduce a network Bayesian information criterion for for selecting the tuning parameter $\lambda_N$, which guarantees the selection consistency of nodal homophily and nodal effects.


\subsection{Asymptotic Theory}

Let's denote $a_w = \max\{\lambda_N w_k: k\in\mathcal{A}\}$ and $b_w=\min\{\lambda_Nw_k: k\in\mathcal{A}^c\}$. When we obtain the regularized estimator $\hat{\theta}$ in \eqref{S2:eq:RMLE}, $a_w$ controls the largest penalty on the true non-zero coefficient of $\theta$, while $b_w$ controls the smallest penalty on the zero-valued coefficient of $\theta$. 

Based on the reparameterization above, we then establish the asymptotic properties of the regularized estimator given below:

\begin{theorem}
\label{Thm1}
    Under Conditions (C1) and (C2) in Appendix, when $ a_w/\sqrt{N} \xrightarrow{\mathrm{P}}  0$, the regularized maximum likelihood estimator $\hat{\theta}_{RMLE}$ of model \eqref{S3:multinominal:Prob} is $\sqrt{N}$-consistent.
\end{theorem}

\begin{theorem}[Oracle property]
\label{Thm2}
    Under Conditions (C1) and (C2) in Appendix, we obtain the following results for the $\sqrt{N}$-consistent estimator $\hat{\theta} = \arg\min_{\theta\in\Theta} Q(\theta)$. 
    \begin{enumerate}
        \item (Selection consistency) If $ a_w / \sqrt{N} \xrightarrow{\mathrm{P}} 0$, then $$\mathrm{P}(\hat{\theta}_{\mathcal{A}^c}=0) \to 1 $$ 
        
        \item (Asymptotic normality) If $a_w/\sqrt{N} \xrightarrow{\mathrm{P}} 0$, and $ b_w/\sqrt{N} \xrightarrow{\mathrm{P}} \infty$, then 
        $$ \sqrt{N}(\hat{\theta}_{\mathcal{A}} - \theta_{0\mathcal{A}}) \xrightarrow{d} \mathcal{N} \left(0, I_\mathcal{A}(\theta_0)^{-1}\right) $$
    \end{enumerate}
\end{theorem}
 
The Theorem \ref{Thm1} implies that the regularized estimator $\hat{\theta}$ is $\sqrt{N}$-consistent. Together with the conclusion in Theorem \ref{Thm2} that the coefficients for irrelevant nodal effects are zero with probability tending to $1$, $\hat{\theta}$ in \eqref{S2:eq:RMLE} can identify the true nodal homophily and nodal effects consistently, i.e., the variable selection consistency. 

The second part of \ref{Thm2} suggests that the estimated coefficients that corresponding to the relevant nodal effects are asymptotically normal, which allows for further statistical inference such as hypothesis testing and confidence interval construction. Together with the selection consistency, the proposed estimator $\hat{\theta}$ enjoys the oracle properties such that the coefficients of true nodal homophily and nodal effects can be estimated equivalently with the case when the true model is known in advance in term of the efficiency.

The condition $a_w/\sqrt{N} \xrightarrow{\mathrm{P}} 0$ indicates that the penalty on the nonzero coefficients of $\theta$ needs to be small, and the condition $b_w/\sqrt{N} \xrightarrow{\mathrm{P}} \infty$ shows that the penalty on the zero coefficients of $\theta_0$ must be large enough to have variable selection consistency. As we use $ 1 / |\hat{\theta}_{MLE}|^{\gamma}$ as the adaptive weights, where the initial estimate is $\sqrt{N}$-consistent by Lemma 1, then $w_j=O_p(1)$ for $j\in\mathcal{A}$ and $w_j \asymp N^{\gamma/2}$ for all $j\in\mathcal{A}^c$. In practice, we set $\gamma=1$ and select $\lambda_N$ based on the newly proposed network Bayesian information criterion, which is introduced as follows.

\subsection{Network Bayesian Information Criterion} 

Although the traditional model selection criteria, such as the Akaike information criterion (AIC, \cite{akaike1998information}) and the Bayesian information criterion (BIC, \cite{schwarz1978estimating}), are commonly used in selecting the regularization parameter $\lambda_N$, the AIC and generalized cross-validation (GCV, \cite{golub1979generalized}) method cannot ensure the consistent selection \citep{wang2007unified, ZhangCH2010AOS}. To address this limitation, we extend the BIC-type criterion and introduce the following network Bayesian information criterion (NBIC) for selecting $\lambda$:
\begin{equation}
\mathrm{NBIC}_\lambda = - \frac{1}{\sqrt{N}} \mathcal{L} (\hat{\theta}_{\lambda}) + \left(\log N \right) \times \text{DF}_\lambda.
\label{S3:BICdef}
\end{equation}
where $\mathcal{L} (\hat{\theta}_{\lambda})$ is the likelihood evaluated at the estimator $\hat{\theta}_{\lambda}$ in \eqref{S2:eq:RMLE} based on tuning parameter $\lambda$, $\text{DF}_\lambda$ denote the number of nonzero elements in $\hat{\theta}_{\lambda} \in \mathbb{R}^p$.  

If two model have the same level of goodness of fit, we prefer the smaller model with smaller $\mathrm{DF}_\lambda$. Thus, the optimal regularization parameter $\lambda$ can then be selected by minimizing NBIC $\hat{\lambda} = \arg \min_\lambda \mathrm{NBIC}_\lambda$. Similar to the classical BIC, the goodness of fit term in \eqref{S3:BICdef} is determined by the likelihood scaled by $1/ \sqrt{N}$, and the model complexity is quantified by $\text{DF}_\lambda$. 

To establish the theoretical guarantee for the proposed BIC-type criterion, we denote the true subset of nodal homophily and effects as $\mathcal{A}=\{k: \theta_k\neq 0\}$, and denote the selected nodal homophily and effects as $\hat{\mathcal{A}}_\lambda=\{k: \hat{\theta}_k\neq 0\}$. We can split $\lambda$ into three mutual exclusive sets $ \Omega_0  = \{ \lambda \in \mathbb{R} : \hat{\mathcal{A}}_{\lambda} = \mathcal{A} \}$, $\Omega_-  = \{ \lambda \in \mathbb{R} : \hat{\mathcal{A}}_{\lambda} \not\supset \mathcal{A} \}$, $\Omega_+  = \{ \lambda \in \mathbb{R} : \hat{\mathcal{A}}_{\lambda} \supset \mathcal{A} \text{ and } \hat{\mathcal{A}}_{\lambda} \neq \mathcal{A}\}$. Let $\asymp $ represents the asymptotic equivalence, where \( f(N) \asymp g(N) \) means that there exist universal constants \( c_1, c_2 > 0 \) such that \( c_1 |g(N)| \leq |f(N)| \leq c_2 |g(N)| \) holds for all sufficiently large \( N \). 
 
\begin{theorem}
\label{thm3}
    Let the adaptive weights $w_k=1/|\hat{\theta}_{k,MLE}|$. Then for any chosen reference regularization parameter $\lambda_N=\lambda_{N}^\circ$ satisfying $\lambda_{N}^\circ \asymp \log N$, we have 
\begin{equation}  
    P\left( \inf_{\lambda \in \Omega_- \cup \Omega_+} \mathrm{NBIC}_\lambda > \mathrm{NBIC}_{\lambda_{N}^\circ} \right) \to 1.
\end{equation}         
as $N \to \infty$. 
\end{theorem}

One can easily show that the reference regularization parameter $\lambda_N = \lambda_{N}^\circ \asymp \log N$ guarantees both two conditions $ a_w/\sqrt{N} \xrightarrow{\mathrm{P}} 0$ and $ b_w /\sqrt{N}\xrightarrow{\mathrm{P}} \infty$ in Theorem \ref{Thm2} with $\gamma=1$ and $\tilde{\theta}=\hat{\theta}_{MLE}$, which is included in Remark \ref{S4:rmk:Tuning} of Supplements Section \ref{Prof.Thm3}. Hence, the $\lambda_N$ could ensures the variable selection consistency in Theorem \ref{Thm2}, meaning that the selected subset of nodal homophily and effects $\hat{\mathcal{A}}=\mathcal{A}$ with probability tending to 1. Thus we refer to $\lambda_N$ as the reference regularization parameter. Meanwhile, the Theorem \ref{thm3} indicates that if the $\lambda$ falls with in the union of $\Omega_-$ and $ \Omega_+$, where the estimator $\hat{\theta}_\lambda$ fails to identify the true nodal homophily and effects $\mathcal{A}$, the corresponding $\mathrm{NBIC}_{\lambda}$ will be greater than $\mathrm{NBIC}_{\lambda_N}$ with probability tending to 1. By choosing the $\hat{\lambda}=\arg\min_{\lambda}\mathrm{NBIC}_{\lambda}$, we have $\mathrm{NBIC}_{\hat{\lambda}}\leq \mathrm{NBIC}_{\lambda_N}$. Therefore, the only possibility is that $\hat{\lambda}$ falls within $\Omega_0$, which ensures variable selection consistency.

Based on the proposition \ref{lemma1} and definition in \eqref{S3:Likelihood}, \(\mathcal{L}(\theta)\) evaluated at any \(\sqrt{N}\)-consistent estimator \(\hat{\theta}\) is of order \(O_{p}(N^{1/2})\). As a consequence, we scale \(\mathcal{L}(\theta) \) by \( 1/\sqrt{N}\) in NBIC \eqref{S3:BICdef}. Note that the $N=n(n-1)/2$ is the total number of dyads included in one network with $n$ nodes, this scaling is different from the classic literature on BIC regularization parameter selection \citep{wang2009shrinkage, jones2003assessment} but crucial for establishing the variable selection consistency result in Theorem \ref{thm3}; see the proof of Theorem \ref{thm3} in Section \ref{Prof.Thm3} of the supplementary material.

\section{Simulation}
\label{S4}

In this section, we evaluate the numerical performance of the regularized maximum likelihood estimator $\hat{\theta}$ in \eqref{S2:eq:RMLE}, and compare it with the non-regularized maximum likelihood estimator $\hat{\theta}_{MLE}$ of \eqref{S3:multinominal:Prob}. 

In our simulation, the dyads of network with $n$ nodes are generated from model \eqref{S3:multinominal:Prob}. We set $m=4$ and include the nodal features $V_1, V_2 \sim \mathcal{N}(0,1)$ and $V_3,V_4 \sim Bernulli(0.5)$. The Gaussian homophily measurements and Chaldean norm are used for $w(\cdot)$ and $g(\cdot)$. To avoid generating full networks or empty network, we set the true coefficients $\theta_0=( 6, -6, 6,  0 , 0 , 0, -6 , 0,  0,  0, 6 , 0 , 0 , 0)^\top$, then the true subset of nodal homophily and nodal effects are $\mathcal{A}=\{1,2,3,7,11\}$. 

To simulate the cases of large-scale networks, we consider $n= 50, 100, 150$ and $200$. The tuning parameter $\lambda$ is chosen based on the BIC-type criterion \eqref{S3:BICdef} from Theorem \ref{thm3}. 
For each of the simulation settings, we obtain $r = 1, \cdots, 1000$ replications $\hat{\theta}^{(r)}=(\hat{\theta}_1^{(r)},\cdots,\hat{\theta}_{14}^{(r)})^\top$ of estimates $\hat{\theta}=(\hat{\theta} ,\cdots,\hat{\theta}_{14} )^\top$ of parameter $\theta=(\theta_1,\cdots,\theta_{14})^\top$. Using these replications, we could evaluate the performance of $\hat{\theta}$ in estimation performance and the  selection of nodal homophily and nodal effects. 

To evaluate the estimation performance, we consider the five measures as below: (i) Empirical Bias (BIAS) $1000^{-1}\sum_{r} (\hat{\theta}^{(r)}_k-\theta_{0,k})$; (ii) Standard Deviation (SD) $(1000^{-1} \sum_{r}  (\hat{\theta}^{(r)}_k-1000^{-1}\sum_{r} \hat{\theta}^{(r)}_k)^2)^{1/2}$; (iii) Root Mean Squared Error (RMSE) $(1000^{-1} \sum_{r}  (\hat{\theta}^{(r)}_k-\theta_0)^2)^{1/2} = (\mathrm{BIAS}^2+ \mathrm{SD}^2)^{1/2}$; (iv) Asymptotic Standard Error (ASE) $1000^{-1} \sum_r \mathrm{SE}_k^{(r)}$, where $\mathrm{SE}_k^{(r)}$ is the standard error $\mathrm{SE}_k$ in $r$-th replication which is obtained from the diagonal elements $[\hat{I}^{(r)}_\mathcal{A}(\hat{\theta})^{-1}]_{kk}$ based on Theorem \ref{Thm2}; (v) Empirical Coverage Probability (CP) for 95\% confidence interval constructed from $\hat{\theta}_k$ and its corresponding asymptotic distribution. Specifically, $$ \mathrm{CP} = 1000^{-1} I\left(\theta_{0k}\in [\hat{\theta}_k^{(r)} -1.96\times\mathrm{SE}_k^{(r)}, \hat{\theta}_k^{(r)} +1.96\times\mathrm{SE}_k^{(r)}]\right),$$ where $I(\cdot)$ is the indicator function. 
    
To evaluate the performance of $\hat{\theta}$ in selection of nodal homophily and nodal effects, we denote the selected set of nodal effects in $r$-th repeat as $\hat{\mathcal{A}}^{(r)}$, $|\hat{\mathcal{A}}^{(r)} |$ represents the number of elements in the set. Then we consider the following four measures of $\hat{\mathcal{A}}$: (i) the average Correct Fit (CF) of selected nodal homophily and effects: $1000^{-1} \sum_{r} I\left(  \hat{\mathcal{A}}^{(r)} = \mathcal{A} \right) $; (ii) the average True Positive Rate (TPR) selected nodal homophily and effects: $1000^{-1} \sum_{r} \frac{ |\hat{\mathcal{A}}^{(r)} \cap \mathcal{A}  |}{|
\mathcal{A}|}$; (iii) the average False Positive Rate (FPR) of selected nodal homophily and effects: $1000^{-1} \sum_{r}\frac{ |\hat{\mathcal{A}}^{(r)} \cap \mathcal{A}^c  |}{|
\mathcal{A}^c|}$; (iv) the average selected Model Size (MS) $1000^{-1} \sum_{r} |\hat{\mathcal{A}}^{(r)}|$. 

\begin{table}[ht]
\centering
\caption{The five measures used to assess the estimation performance}
\label{SimTable1}
\footnotesize
\begin{tabular}{@{}ccccccc@{}}
\toprule
\multicolumn{2}{c}{}          & $\theta_1$ & $\theta_2$ & $\theta_3$ & $\theta_7$ & $\theta_{11}$ \\ \midrule
\multirow{5}{*}{n=50}  & BIAS & -0.2008  & 0.0504   & -0.1025  & 0.0313   & -0.0432   \\
                       & SD   & 1.0956   & 0.5145   & 0.6740   & 0.5962   & 0.5463    \\
                       & RMSE & 1.1139   & 0.5170   & 0.6818   & 0.5970   & 0.5480    \\
                       & ASE  & 4.7917   & 0.5058   & 0.6005   & 0.5384   & 0.5379    \\
                       & CP   & 0.9250   & 0.9580   & 0.9120   & 0.9490   & 0.9530    \\ \midrule
\multirow{5}{*}{n=100} & BIAS & -0.0762  & 0.0448   & -0.0669  & 0.0368   & -0.0422   \\
                       & SD   & 0.3352   & 0.2120   & 0.2914   & 0.2239   & 0.2240    \\
                       & RMSE & 0.3438   & 0.2166   & 0.2990   & 0.2269   & 0.2280    \\
                       & ASE  & 0.3341   & 0.2389   & 0.2898   & 0.2427   & 0.2426    \\
                       & CP   & 0.9420   & 0.9610   & 0.9380   & 0.9610   & 0.9610    \\ \midrule
\multirow{5}{*}{n=200} & BIAS & -0.0169  & 0.0105   & -0.0175  & 0.0087   & -0.0113   \\
                       & SD   & 0.1702   & 0.1085   & 0.1452   & 0.1135   & 0.1126    \\
                       & RMSE & 0.1710   & 0.1090   & 0.1462   & 0.1139   & 0.1131    \\
                       & ASE  & 0.1659   & 0.1174   & 0.1433   & 0.1181   & 0.1180    \\
                       & CP   & 0.9470   & 0.9680   & 0.9500   & 0.9540   & 0.9560    \\ \midrule
\multirow{5}{*}{n=400} & BIAS & -0.0071  & 0.0034   & -0.0039  & 0.0033   & -0.0029   \\
                       & SD   & 0.0812   & 0.0532   & 0.0707   & 0.0554   & 0.0559    \\
                       & RMSE & 0.0815   & 0.0533   & 0.0708   & 0.0555   & 0.0559    \\
                       & ASE  & 0.0822   & 0.0581   & 0.0711   & 0.0582   & 0.0581    \\
                       & CP   & 0.9530   & 0.9750   & 0.9520   & 0.9630   & 0.9600    \\ \bottomrule
\end{tabular}
\end{table}

Table \ref{SimTable1} presents the estimation performance of proposed $\hat{\theta}_{RMLE}$ and the comparison with the non-regularized estimator $\hat{\theta}_{MLE}$. As the sample size (size of network) increases, the bias of $\hat{\theta}$ decreases in all simulation settings, which demonstrates the the consistency in Theorem \ref{Thm2} and Lemma \ref{lemma1}. For any given sample size, the regularized estimator $\hat{\theta}$ could achieve a slightly larger bias and smaller RMSE compared to the $\hat{\theta}_{MLE}$. The CP is close to 95\% in all cases, which further validates the inferential applicability f the asymptotic normality in Theorem \ref{Thm2}.  


Table \ref{SimTable2} reports the four measures used to assess the selection performance for the nodal homophily and nodal effects. From the results, the proposed Regularized MLE achieve $\mathrm{TPR}=1$ in all cases. And  as $N$ increases, we observed a increasing average Correct Fit and decreasing False Discover Rate result, the average model size selected converge to $|\mathcal{A}|=5$ with decreasing standard deviations, which indicates the selection consistency achieved by  $\mathrm{NBIC}_\lambda$.  

\begin{table}[ht]
\centering
\caption{The four measures used to assess the variable selection performance of regularized maximum likelihood estimator $\hat{\theta}$, and the corresponding standard deviations are presented in parentheses}  
\label{SimTable2}
\begin{tabular}{@{}ccccc@{}}
\toprule
n   & CF              & TPR   & FPR             & MZ             \\ \midrule
50  & 0.8948 (0.0800)   & 1.0000 (0.0000) & 0.1637 (0.1245) & 6.473 (1.1201) \\
100 & 0.9256 (0.0726) & 1.0000 (0.0000) & 0.1158 (0.1129) & 6.042 (1.0165) \\
200 & 0.9446 (0.0576) & 1.0000 (0.0000) & 0.0861 (0.0897) & 5.775 (0.8069) \\
400 & 0.9541 (0.0563) & 1.0000 (0.0000) & 0.0714 (0.0875) & 5.643 (0.7875) \\ \bottomrule
\end{tabular}
\end{table}

\section{Real Data Analysis}
\label{S5}

In this section, we apply the proposed method to a publicly available dataset\footnote{\url{https://snap.stanford.edu/data/gemsec-Deezer.html}} from Deezer, a widely used music streaming platform in Europe. The data were collected in November 2017 and documented by \citet{rozemberczki2019gemsec}. It contains both individual-level features and the directed follower-followee relationships among users, which together form a directed social network. Each user’s musical preferences across 84 distinct genres are recorded as nodal features. Our objective is to identify the influential nodal homophily and nodal effect terms associated with musical genre preferences based on the observed network structure among Deezer users.
             
Specifically, the dataset comprises a total of $n=143,884$ users from three European countries, along with their self-reported musical preferences across 84 genres. These preferences are represented as 84 binary (dummy) variables. Detailed descriptions of the genre features are provided in Table \ref{S5EAgenres} in the supplementary materials (Section \ref{SuprealdataAnalysis}).
To ensure algorithmic stability and compliance with the theoretical conditions outlined in Section \ref{S3}, we preprocess the data by removing isolated nodes (i.e., users without any incoming or outgoing edges) and those with missing values in the nodal features. We then conduct our analysis on a selected subnetwork comprising 500 nodes. The estimation results for the selected nodal homophily terms and nodal effects are reported in Table \ref{S5:EAre}, ordered by magnitude. In the table, the selected nodal homophily effects, out-degree nodal effects, and in-degree nodal effects are highlighted in blue, orange, and green, respectively.

When focusing on the strongest positive nodal homophily effects, we observe that shared preferences for \textit{Chicago Blues} and \textit{Classical} music are significantly associated with the formation of ties. 
\textit{Chicago blues} , which emerged in the early twentieth century, is characterized by its urban roots and its emphasis on expressive improvisation. As noted by \citep{grazian2004symbolic}, the genre holds substantial cultural significance, having historically provided a communal and cathartic space for marginalized groups. Its themes of struggle and resilience may resonate with listeners who share similar socio-cultural values, thus facilitating social connections. The selected nodal homophily in \textit{Chicago blues} may reflect a symbolic affinity formed through common musical taste, which serves as a proxy for shared identity, emotional orientation, or cultural capital. 

\textit{Classical} music, rooted in European traditions spanning Baroque to Romantic eras, emphasizes structure, complexity, and emotional nuance. Preference for Classical music has been associated with higher levels of educational attainment and intellectual orientation, and may serve as an indicator of cultural capital and aesthetic sophistication \citep{wang2016classical, Tekman01102002}. Thus, shared interest in Classical music may reflect alignment in values such as discipline, introspection, and historical consciousness, which is consistent with the strong positive nodal homophily and in-edge nodal effects estimated in our network analysis.
A similar pattern is observed for \textit{Asian Music}, which is characterized by strong regional attributes and often embodies a distinct sense of cultural identity \citep{matsue2013stars}. The structural and aesthetic contrasts between Asian and Western musical traditions may give rise to asymmetrical but meaningful cross-cultural connections, potentially facilitating social ties based on curiosity or cultural appreciation. For example, the transnational popularity of genres such as Japanese \textit{city pop} from the 1960s to early 1990s and the hybridized appeal of \textit{K-pop} since the 1990s illustrate the role of Asian music as a medium of intercultural engagement \citep{fuhr2015globalization}.

\begin{table}[ht] 
\caption{The selected non-zero \textcolor{blue}{nodal homophily}, \textcolor{orange!80!black}{nodal out-edge effects} and nodal \textcolor{green!50!black}{in-edge effect}} 
\label{S5:EAre} 
\centering
\begin{tabular}{@{}cc|cc@{}}
    \toprule
    \multirow{2}{*}{\textbf{Selected Effects}} & \multirow{2}{*}{\textbf{Coefficients}} & \multirow{2}{*}{\textbf{Selected Effects}} & \multirow{2}{*}{\textbf{Coefficients}} \\
                                               &                                        &                                            &                                        \\ \midrule
    \textcolor{blue}{Chicago Blues}            & 40.2569                                & \textcolor{orange!80!black}{Dance}                  & -1.0290                                \\
    \textcolor{green!50!black}{Classical}               & 34.1537                                & \textcolor{green!50!black}{Soundtracks}             & -2.1634                                \\
    \textcolor{blue}{Classical}                & 30.9461                                & \textcolor{orange!80!black}{Rap/Hip Hop}            & -2.2473                                \\
    \textcolor{blue}{Asian Music}              & 17.9721                                & \textcolor{orange!80!black}{Pop}                    & -5.1447                                \\
    \textcolor{green!50!black}{Asian Music}             & 11.6706                                & \textcolor{green!50!black}{Country}                 & -8.3412                                \\
    \textcolor{green!50!black}{Metal}                   & 11.2522                                & \textcolor{green!50!black}{Rock \& Roll/Rockabilly} & -11.8629                               \\
    \textcolor{blue}{Soul \& Funk}             & 1.3104                                 & \textcolor{green!50!black}{Contemporary Soul}       & -15.8325                               \\
    \textcolor{green!50!black}{East Coast}              & 0.9920                                 & \textcolor{blue}{TV Soundtracks}           & -19.7623                               \\
    \textcolor{green!50!black}{International Pop}       & 0.3379                                 & \textcolor{green!50!black}{Hard Rock}               & -24.8668                               \\
    \textcolor{blue}{Pop}                      & 0.0415                                 & \textcolor{green!50!black}{TV Soundtracks}                                    & -37.7950                               \\
    \textcolor{blue}{Dubstep}                  & -0.1913                                &  \textcolor{green!50!black}{Trance}          &   -49.0946                                \\
    \textcolor{orange!80!black}{Electro}                & -0.6628                                &                  &                             \\ 
    \bottomrule
    \end{tabular}
\end{table}

\textit{Trance} music is defined by repetitive rhythmic patterns, gradual melodic build-ups, and climactic drops, emphasizing immersive auditory experiences and sustained energy. While these stylistic elements are well-suited to dance-oriented contexts, they may be perceived by non-electronic music audiences as lacking in structural variation or narrative progression. The genre’s reliance on synthesized timbres and abstract sonic textures, rather than lyric-driven content, can further limit emotional accessibility for listeners accustomed to more narrative or affect-laden musical forms such as pop or blues. 
Moreover, \textit{Trance} culture is often embedded within electronic music subcultures, where social interactions are concentrated in niche settings such as festivals and underground venues. This subcultural orientation may contribute to relatively insular social connectivity patterns, potentially inhibiting cross-group social integration within broader networks. 
Similarly, \textit{TV soundtracks} represent a heterogeneous musical category whose stylistic diversity and auxiliary narrative function complicate the emergence of cohesive listening communities. Their aesthetic appeal is often contingent on viewers' prior familiarity with the associated television or cinematic works, which restricts broader musical engagement beyond these media contexts. Although such soundtracks can foster strong group identity among fans of specific series, their lack of a unified musical identity limits their capacity to serve as stable bases for social tie formation in general network settings.
These findings align with our empirical results presented in Table \ref{S5:EAre}, where both \textit{Trance} and \textit{TV soundtracks} are associated with negative nodal homophily and i-edge effects. The observed patterns appear consistent with the socio-cultural and affective characteristics of these genres, which may reduce their potential to facilitate new social connections in social networks.


\section{Conclusion}
\label{S6}
In this article, we propose a novel random directed network model with nodal homophily and nodal effects, and also a regularized maximum likelihood estimator to simultaneously perform the estimation and selection of nodal homophily and nodal effects. The proposed estimator is shown to achieve $\sqrt{N}$-consistency and possess the oracle properties, including selection consistency and asymptotic normality. To facilitate practical implementation, we introduce a network Bayesian information criterion for selecting the regularization parameter, which further guarantees consistent selection of relevant nodal covariates. 
Through extensive numerical studies, we demonstrate that the proposed methods can estimate the true coefficient consistently and achieve the selection consistency, validating both the theoretical asymptotic properties and finite-sample performance.   Furthermore, an empirical application to a real-world social network dataset illustrates the practical utility of our approach in identifying meaningful nodal homophily and nodal effects in large-scale directed networks.
Overall, the proposed regularization and selection approach for nodal homophily and nodal effects contributes to the network analysis by enabling interpretable and scalable inference in directed network with complex nodal information. It provides feasible and powerful tools with theoretic guarantee for empirical applications of network science.


\newpage





\renewcommand{\theequation}{A.\arabic{equation}}
\setcounter{equation}{0}
\renewcommand{\thesubsection}{A.\arabic{subsection}}
\renewcommand{\thetheorem}{A.\arabic{theorem}}
\setcounter{theorem}{0}
 
 \setstretch{1}  
 \setlength{\bibsep}{4pt plus 0.3ex} 

\bibliographystyle{chicago} 
\bibliography{reference} 

\begin{thebibliography}{}

\bibitem[\protect\citeauthoryear{Akaike}{Akaike}{1998}]{akaike1998information}
Akaike, H. (1998).
\newblock Information theory and an extension of the maximum likelihood
  principle.
\newblock In {\em Selected papers of hirotugu akaike}, pp.\  199--213.
  Springer.

\bibitem[\protect\citeauthoryear{Albert and Barab{\'a}si}{Albert and
  Barab{\'a}si}{2002}]{albert2002statistical}
Albert, R. and A.-L. Barab{\'a}si (2002).
\newblock Statistical mechanics of complex networks.
\newblock {\em Reviews of Modern Physics\/}~{\em 74\/}(1), 47.

\bibitem[\protect\citeauthoryear{Bramoull{\'e}, Djebbari, and
  Fortin}{Bramoull{\'e} et~al.}{2020}]{bramoulle2020peer}
Bramoull{\'e}, Y., H.~Djebbari, and B.~Fortin (2020).
\newblock Peer effects in networks: A survey.
\newblock {\em Annual Review of Economics\/}~{\em 12\/}(1), 603--629.

\bibitem[\protect\citeauthoryear{Chen, Song, Bai, Lin, and Chang}{Chen
  et~al.}{2010}]{chen2010parallel}
Chen, W.-Y., Y.~Song, H.~Bai, C.-J. Lin, and E.~Y. Chang (2010).
\newblock Parallel spectral clustering in distributed systems.
\newblock {\em IEEE Transactions on Pattern Analysis and Machine
  Intelligence\/}~{\em 33\/}(3), 568--586.

\bibitem[\protect\citeauthoryear{De~Paula}{De~Paula}{2020}]{DePaula2020}
De~Paula, {\'A}. (2020).
\newblock Econometric models of network formation.
\newblock {\em Annual Review of Economics\/}~{\em 12\/}(1), 775--799.

\bibitem[\protect\citeauthoryear{Erd\H{o}s and R\H{e}nyi}{Erd\H{o}s and
  R\H{e}nyi}{1959}]{ERnetwork1959}
Erd\H{o}s, P. and A.~R\H{e}nyi (1959).
\newblock On random graphs. {I}.
\newblock {\em Publ. Math. Debrecen\/}~{\em 6}, 290--297.

\bibitem[\protect\citeauthoryear{Fan and Lv}{Fan and Lv}{2008}]{fan2008sure}
Fan, J. and J.~Lv (2008).
\newblock Sure independence screening for ultrahigh dimensional feature space.
\newblock {\em Journal of the Royal Statistical Society Series B: Statistical
  Methodology\/}~{\em 70\/}(5), 849--911.

\bibitem[\protect\citeauthoryear{Fuhr}{Fuhr}{2015}]{fuhr2015globalization}
Fuhr, M. (2015).
\newblock {\em Globalization and popular music in South Korea: Sounding out
  K-pop}.
\newblock Routledge.

\bibitem[\protect\citeauthoryear{Gao, Liu, Pan, and Wang}{Gao
  et~al.}{2024}]{gao2024citation}
Gao, T., J.~Liu, R.~Pan, and H.~Wang (2024).
\newblock Citation counts prediction of statistical publications based on
  multi-layer academic networks via neural network model.
\newblock {\em Expert Systems with Applications\/}~{\em 238}, 121634.

\bibitem[\protect\citeauthoryear{Golub, Heath, and Wahba}{Golub
  et~al.}{1979}]{golub1979generalized}
Golub, G.~H., M.~Heath, and G.~Wahba (1979).
\newblock Generalized cross-validation as a method for choosing a good ridge
  parameter.
\newblock {\em Technometrics\/}~{\em 21\/}(2), 215--223.

\bibitem[\protect\citeauthoryear{Graham}{Graham}{2017}]{graham2017econometric}
Graham, B.~S. (2017).
\newblock An econometric model of network formation with degree heterogeneity.
\newblock {\em Econometrica\/}~{\em 85\/}(4), 1033--1063.

\bibitem[\protect\citeauthoryear{Grazian}{Grazian}{2004}]{grazian2004symbolic}
Grazian, D. (2004).
\newblock The symbolic economy of authenticity in the chicago blues scene.
\newblock {\em Music scenes: Local, Translocal, and Virtual\/}, 31--47.

\bibitem[\protect\citeauthoryear{Hastie, Tibshirani, and Wainwright}{Hastie
  et~al.}{2015}]{hastie2015statistical}
Hastie, T., R.~Tibshirani, and M.~Wainwright (2015).
\newblock {\em Statistical Learning with Sparsity: The Lasso and
  Generalizations}.
\newblock CRC Press.

\bibitem[\protect\citeauthoryear{Hoff}{Hoff}{2005}]{hoff2005bilinear}
Hoff, P.~D. (2005).
\newblock Bilinear mixed-effects models for dyadic data.
\newblock {\em Journal of the American Statistical Association\/}~{\em
  100\/}(469), 286--295.

\bibitem[\protect\citeauthoryear{Hunter}{Hunter}{2007}]{Hunter2007}
Hunter, D.~R. (2007).
\newblock Curved exponential family models for social networks.
\newblock {\em Social Networks\/}~{\em 29\/}(2), 216--230.

\bibitem[\protect\citeauthoryear{Hunter, Handcock, Butts, Goodreau, and
  Morris}{Hunter et~al.}{2008}]{hunter2008ergm}
Hunter, D.~R., M.~S. Handcock, C.~T. Butts, S.~M. Goodreau, and M.~Morris
  (2008).
\newblock ergm: A package to fit, simulate and diagnose exponential-family
  models for networks.
\newblock {\em Journal of Statistical Software\/}~{\em 24\/}(3), nihpa54860.

\bibitem[\protect\citeauthoryear{Jiang and Chen}{Jiang and
  Chen}{2015}]{jiang2015natergm}
Jiang, S. and H.~Chen (2015).
\newblock Natergm: A model for examining the role of nodal attributes in
  dynamic social media networks.
\newblock {\em IEEE Transactions on Knowledge and Data Engineering\/}~{\em
  28\/}(3), 729--740.

\bibitem[\protect\citeauthoryear{Johnson and Wichern}{Johnson and
  Wichern}{1992}]{johnson1992applied}
Johnson, R.~A. and D.~W. Wichern (1992).
\newblock {\em Applied multivariate statistical analysis}.
\newblock Prentice Hall.

\bibitem[\protect\citeauthoryear{Jones and Handcock}{Jones and
  Handcock}{2003}]{jones2003assessment}
Jones, J.~H. and M.~S. Handcock (2003).
\newblock An assessment of preferential attachment as a mechanism for human
  sexual network formation.
\newblock {\em Proceedings of the Royal Society of London. Series B: Biological
  Sciences\/}~{\em 270\/}(1520), 1123--1128.

\bibitem[\protect\citeauthoryear{Lehmann, Lautrup, and Jackson}{Lehmann
  et~al.}{2003}]{lehmann2003citation}
Lehmann, S., B.~Lautrup, and A.~D. Jackson (2003).
\newblock Citation networks in high energy physics.
\newblock {\em Physical Review E\/}~{\em 68\/}(2), 026113.

\bibitem[\protect\citeauthoryear{Lei and Rinaldo}{Lei and
  Rinaldo}{2015}]{LeiRinaldo}
Lei, J. and A.~Rinaldo (2015).
\newblock Consistency of spectral clustering in stochastic block models.
\newblock {\em The Annals of Statistics\/}, 215--237.

\bibitem[\protect\citeauthoryear{Li, Zhong, and Zhu}{Li
  et~al.}{2012}]{LiZhong2012feature}
Li, R., W.~Zhong, and L.~Zhu (2012).
\newblock Feature screening via distance correlation learning.
\newblock {\em Journal of the American Statistical Association\/}~{\em
  107\/}(499), 1129--1139.

\bibitem[\protect\citeauthoryear{Matsue}{Matsue}{2013}]{matsue2013stars}
Matsue, J.~M. (2013).
\newblock Stars to the state and beyond: Globalization, identity, and asian
  popular music.
\newblock {\em The Journal of Asian Studies\/}~{\em 72\/}(1), 5--20.

\bibitem[\protect\citeauthoryear{McMillan}{McMillan}{2022}]{McMillan2022}
McMillan, C. (2022).
\newblock Strong and weak tie homophily in adolescent friendship networks: An
  analysis of same-race and same-gender ties.
\newblock {\em Network Science\/}~{\em 10\/}(3), 283--300.

\bibitem[\protect\citeauthoryear{McPherson, Smith-Lovin, and Cook}{McPherson
  et~al.}{2001}]{McPherson2001}
McPherson, M., L.~Smith-Lovin, and J.~M. Cook (2001).
\newblock Birds of a feather: Homophily in social networks.
\newblock {\em Annual Review of Sociology\/}~{\em 27\/}(1), 415--444.

\bibitem[\protect\citeauthoryear{Neal}{Neal}{2003}]{Neal2003}
Neal, P. (2003).
\newblock Sir epidemics on a bernoulli random graph.
\newblock {\em Journal of Applied Probability\/}~{\em 40\/}(3), 779--782.

\bibitem[\protect\citeauthoryear{Newman}{Newman}{2001}]{newman2001scientific}
Newman, M.~E. (2001).
\newblock Scientific collaboration networks. i. network construction and
  fundamental results.
\newblock {\em Physical Review E\/}~{\em 64\/}(1), 016131.

\bibitem[\protect\citeauthoryear{No{\"e}, Whitaker, and Allen}{No{\"e}
  et~al.}{2016}]{NEO2016}
No{\"e}, N., R.~M. Whitaker, and S.~M. Allen (2016).
\newblock Personality homophily and the local network characteristics of
  facebook.
\newblock In {\em Proceedings of the 2016 IEEE International Conference on
  Advances in Social Networks Analysis and Mining}, pp.\  386--393. IEEE.

\bibitem[\protect\citeauthoryear{Rozemberczki, Davies, Sarkar, and
  Sutton}{Rozemberczki et~al.}{2019}]{rozemberczki2019gemsec}
Rozemberczki, B., R.~Davies, R.~Sarkar, and C.~Sutton (2019).
\newblock Gemsec: Graph embedding with self clustering.
\newblock In {\em Proceedings of the 2019 IEEE International Conference on
  Advances in Social Networks Analysis and Mining}, pp.\  65--72. ACM.

\bibitem[\protect\citeauthoryear{Schwarz}{Schwarz}{1978}]{schwarz1978estimating}
Schwarz, G. (1978).
\newblock {Estimating the Dimension of a Model}.
\newblock {\em The Annals of Statistics\/}~{\em 6\/}(2), 461 -- 464.

\bibitem[\protect\citeauthoryear{Stewart, Schweinberger, Bojanowski, and
  Morris}{Stewart et~al.}{2019}]{Stewart2019}
Stewart, J., M.~Schweinberger, M.~Bojanowski, and M.~Morris (2019).
\newblock Multilevel network data facilitate statistical inference for curved
  ergms with geometrically weighted terms.
\newblock {\em Social Networks\/}~{\em 59}, 98--119.

\bibitem[\protect\citeauthoryear{Stillwell and Kosinski}{Stillwell and
  Kosinski}{2015}]{SK2015dataset}
Stillwell, D. and M.~Kosinski (2015).
\newblock mypersonality project website.

\bibitem[\protect\citeauthoryear{Tekman and and}{Tekman and
  and}{2002}]{Tekman01102002}
Tekman, H.~G. and N.~H. and (2002).
\newblock Music and social identity: Stylistic identification as a response to
  musical style.
\newblock {\em International Journal of Psychology\/}~{\em 37\/}(5), 277--285.

\bibitem[\protect\citeauthoryear{Toivonen, Kovanen, Kivel{\"a}, Onnela,
  Saram{\"a}ki, and Kaski}{Toivonen et~al.}{2009}]{Toivonen2009}
Toivonen, R., L.~Kovanen, M.~Kivel{\"a}, J.-P. Onnela, J.~Saram{\"a}ki, and
  K.~Kaski (2009).
\newblock A comparative study of social network models: Network evolution
  models and nodal attribute models.
\newblock {\em Social Networks\/}~{\em 31\/}(4), 240--254.

\bibitem[\protect\citeauthoryear{Wang and Leng}{Wang and
  Leng}{2007}]{wang2007unified}
Wang, H. and C.~Leng (2007).
\newblock Unified lasso estimation by least squares approximation.
\newblock {\em Journal of the American Statistical Association\/}~{\em
  102\/}(479), 1039--1048.

\bibitem[\protect\citeauthoryear{Wang, Li, and Leng}{Wang
  et~al.}{2009}]{wang2009shrinkage}
Wang, H., B.~Li, and C.~Leng (2009).
\newblock Shrinkage tuning parameter selection with a diverging number of
  parameters.
\newblock {\em Journal of the Royal Statistical Society Series B: Statistical
  Methodology\/}~{\em 71\/}(3), 671--683.

\bibitem[\protect\citeauthoryear{Wang}{Wang}{2016}]{wang2016classical}
Wang, J. (2016).
\newblock Classical music: a norm of ``common" culture embedded in cultural
  consumption and cultural diversity.
\newblock {\em International Review of the Aesthetics and Sociology of
  Music\/}, 195--205.

\bibitem[\protect\citeauthoryear{Wang, Fellows, and Handcock}{Wang
  et~al.}{2024}]{Wangetal2024}
Wang, Z., I.~E. Fellows, and M.~S. Handcock (2024).
\newblock Understanding networks with exponential-family random network models.
\newblock {\em Social Networks\/}~{\em 78}, 81--91.

\bibitem[\protect\citeauthoryear{Watts and Strogatz}{Watts and
  Strogatz}{1998}]{Watts1998}
Watts, D.~J. and S.~H. Strogatz (1998).
\newblock Collective dynamics of ``small-world" networks.
\newblock {\em Nature\/}~{\em 393\/}(6684), 440--442.

\bibitem[\protect\citeauthoryear{Wimmer and Lewis}{Wimmer and
  Lewis}{2010}]{Wimmer2010}
Wimmer, A. and K.~Lewis (2010).
\newblock Beyond and below racial homophily: Erg models of a friendship network
  documented on facebook.
\newblock {\em American Journal of Sociology\/}~{\em 116\/}(2), 583--642.

\bibitem[\protect\citeauthoryear{Xing, Tan, Zhong, and Shi}{Xing
  et~al.}{2025}]{XingCalms2025}
Xing, Z., H.~Tan, W.~Zhong, and L.~Shi (2025).
\newblock Calms: Constrained adaptive lasso with multi-directional signals for
  latent networks reconstruction.
\newblock {\em Neurocomputing\/}~{\em 630}, 129545.

\bibitem[\protect\citeauthoryear{Xing, Wan, Wen, and Zhong}{Xing
  et~al.}{2024}]{xing2024golfs}
Xing, Z., Y.~Wan, J.~Wen, and W.~Zhong (2024).
\newblock Golfs: feature selection via combining both global and local
  information for high dimensional clustering.
\newblock {\em Computational Statistics\/}~{\em 39\/}(5), 2651--2675.

\bibitem[\protect\citeauthoryear{Yan, Jiang, Fienberg, and Leng}{Yan
  et~al.}{2019}]{Yan2019JASA}
Yan, T., B.~Jiang, S.~E. Fienberg, and C.~Leng (2019).
\newblock Statistical inference in a directed network model with covariates.
\newblock {\em Journal of the American Statistical Association\/}~{\em
  114\/}(526), 857--868.

\bibitem[\protect\citeauthoryear{Yan, Leng, and Zhu}{Yan
  et~al.}{2016}]{Yan2016AOS}
Yan, T., C.~Leng, and J.~Zhu (2016).
\newblock Asymptotics in directed exponential random graph models with an
  increasing bi-degree sequence.
\newblock {\em The Annals of Statistics\/}~{\em 44\/}(1), 31--57.

\bibitem[\protect\citeauthoryear{Zhang}{Zhang}{2010}]{ZhangCH2010AOS}
Zhang, C.~H. (2010).
\newblock {Nearly unbiased variable selection under minimax concave penalty}.
\newblock {\em The Annals of Statistics\/}~{\em 38\/}(2), 894 -- 942.

\bibitem[\protect\citeauthoryear{Zhao, Levina, and Zhu}{Zhao
  et~al.}{2012}]{Zhao2012}
Zhao, Y., E.~Levina, and J.~Zhu (2012).
\newblock Consistency of community detection in networks under degree-corrected
  stochastic block models.
\newblock {\em The Annals of Statistics\/}~{\em 40\/}(4), 2266--2292.

\bibitem[\protect\citeauthoryear{Zheng, Salganik, and Gelman}{Zheng
  et~al.}{2006a}]{Zheng2006JASA}
Zheng, T., M.~J. Salganik, and A.~Gelman (2006a).
\newblock How many people do you know in prison? using overdispersion in count
  data to estimate social structure in networks.
\newblock {\em Journal of the American Statistical Association\/}~{\em
  101\/}(474), 409--423.

\bibitem[\protect\citeauthoryear{Zheng, Salganik, and Gelman}{Zheng
  et~al.}{2006b}]{zheng2006many}
Zheng, T., M.~J. Salganik, and A.~Gelman (2006b).
\newblock How many people do you know in prison? using overdispersion in count
  data to estimate social structure in networks.
\newblock {\em Journal of the American Statistical Association\/}~{\em
  101\/}(474), 409--423.

\bibitem[\protect\citeauthoryear{Zou}{Zou}{2006}]{zou2006adaptive}
Zou, H. (2006).
\newblock The adaptive lasso and its oracle properties.
\newblock {\em Journal of the American Statistical Association\/}~{\em
  101\/}(476), 1418--1429.

\bibitem[\protect\citeauthoryear{Zou, Lan, Wang, and and}{Zou
  et~al.}{2017}]{Zou02012017}
Zou, T., W.~Lan, H.~Wang, and C.-L.~T. and (2017).
\newblock Covariance regression analysis.
\newblock {\em Journal of the American Statistical Association\/}~{\em
  112\/}(517), 266--281.

\end{thebibliography}

\end{document}